\documentclass[12pt]{iopart}

\usepackage{graphicx}

\begin{document}

\jl{22}

\title{Current rectification in a single molecule diode: the role of electrode coupling}

\author{Siya Sherif$^{1,2}$, Gabino Rubio-Bollinger$^2$, Elena Pinilla-Cienfuegos$^3$,  Eugenio Coronado $^3$,Juan Carlos Cuevas$^4$  and Nicol\'{a}s Agra\"{\i}t$^{1,2}$}

\address{$^1$ Instituto Madrile\~{n}o de Estudios Avanzados en Nanociencia (IMDEA - Nanoscience) , Faraday, 9, Ciudad Universitaria de Cantoblanco,
28049, Madrid, Spain}

\address{$^2$ Departamento de F\'{\i}sica de la Materia Condensada and Condensed Matter Physics Center (IFIMAC), Facultad de Ciencias,
c/ Francisco Tom\'{a}s y Valiente, 7
Universidad Aut\'{o}noma de Madrid
28049 Madrid, Spain}

\address{$^3$ Instituto de Ciencia Molecular (ICMol), Universidad de Valencia, 
Catedr\'{a}tico Jos\'{e} Beltr\'{a}n Mart\'{\i}nez No. 2, 46980 Paterna, Spain}

\address{$^4$  Departamento de F\'{\i}sica Te\'{o}rica de la Materia Condensada and Condensed Matter Physics Center (IFIMAC), Universidad Aut\'{o}noma de Madrid, E-28049 Madrid, Spain}

\ead{nicolas.agrait@uam.es }
\begin{abstract}
We demonstrate large rectification ratios ($>100$) in single-molecule junctions based on a metal-oxide cluster (polyoxometalate), using a scanning tunneling microscope (STM) both at ambient conditions and at low temperature. These rectification ratios are the largest ever observed in a single-molecule junction, and in addition these junctions sustain current densities larger than $10^5$ A/cm$^2$. By following the variation of the I-V characteristics with tip-molecule separation we demonstrate unambiguously that rectification is due to asymmetric coupling to the electrodes of a molecule with an asymmetric level structure. This mechanism can be implemented in other type of molecular junctions using both organic and inorganic molecules and provides a simple strategy for the rational design of molecular diodes.
\end{abstract}


%
\vspace{2pc}
%

%
%
%

\section{Introduction}

Molecular electronics aims at using molecules as active elements in nanoscale electronic circuits and it was born with the prediction that a single molecule could function as a diode \cite{Aviram}. Molecular rectification has been shown in many-molecule devices\cite{Metzger1, Metzger2, Kushmerick, Guedon}, in some cases with rectification ratios up to about 150 \cite{Ashwell, Nijhuis, Nerngchamnong}, but often the current density is very low.  Single-molecule diodes have also been reported \cite{Elbing, Diez, Lortscher, Batra}, but in all cases they suffer from low rectification ratios ($< 10$), and often the underlying physical mechanism remains unclear. Here we show that a single molecule based on a metal-oxide cluster (polyoxometalate) can reliably and reproducibly work as a diode exhibiting rectification ratios larger than several hundred both at room and cryogenic temperatures. Our single-molecule diodes have very high current densities $(>10^5 \ \rm{A/cm}^2)$ and, more importantly, we demonstrate unambiguously that rectification is due to asymmetric coupling to the electrodes of a molecule with an asymmetric level structure. Such a physical mechanism is highly transferable and, in principle, it can be implemented in macroscopic molecular tunneling junctions.

Over the years, different physical mechanisms have been proposed that could potentially lead to electrical current rectification in a molecular junction. Probably the simplest one, conceptually speaking, is that in which a molecule is asymmetrically connected to the electrodes. This may lead to an asymmetric voltage profile across the junction that, in turn, induces an asymmetry in the current-voltage characteristics (I-V). This mechanism has been theoretically discussed by several authors in the context of molecular junctions \cite{Kornilovitch, Larade, Taylor, Cuevas}, and it is believed to be behind many observations of small asymmetries in the I-Vs \cite{Reichert}. However, no functional diode, i.e. with a large rectification ratio, based on this simple mechanism has ever been demonstrated. In principle, an asymmetric metal-molecule-metal coupling can be achieved by employing intrinsically asymmetric molecules, something that can be done by using asymmetric molecular backbones or by introducing two different anchoring groups. This strategy, however, has not been very successful so far. On the other hand, an asymmetry can be created by controlling the relative metal-molecule couplings by, for instance, controlling the metal-molecule separation. It has been argued that although such a procedure may lead to the desired asymmetry, it is necessarily accompanied by a dramatic reduction of the current\cite{Batra}. The goal of this work is to show that an appropriate choice of the molecule together with a precise control of the metal-molecule coupling leads indeed to single-molecule diodes with both huge rectification ratios and high current densities which can operate even at room temperatures and in ambient conditions. 

\section{Experimental Section}

The polyoxometalate cluster $\rm [DyP_{5}W_{30}O_{110}]^{12-}$ used in this work was prepared following the procedure described in ref. 19 \cite{Cardona}. The polyoxometalate molecules were dissolved in milliQ water to obtain dilute solution that was sprayed using a commercial nebulizer (Meinhard TR-30-K1) onto an annealed gold substrate (Arrandee). This deposition method gives a better control on coverage of molecules on the substrate than other wet deposition methods like drop casting, dipping, etc.
We used two different homebuilt scanning tunneling microscopes: a low-drift, high stability STM for the ambient experiments and a low temperature STM attached to an Oxford instruments Heliox $^3$He refrigerator for the low temperature (350 mK) measurements.  I-V characteristics were acquired by ramping the bias voltage while the feedback was suspended. They typically consists of 1024 data points acquired in about 100 ms.

\section{Result and Analysis}

\begin{figure}
\begin{center}
\includegraphics{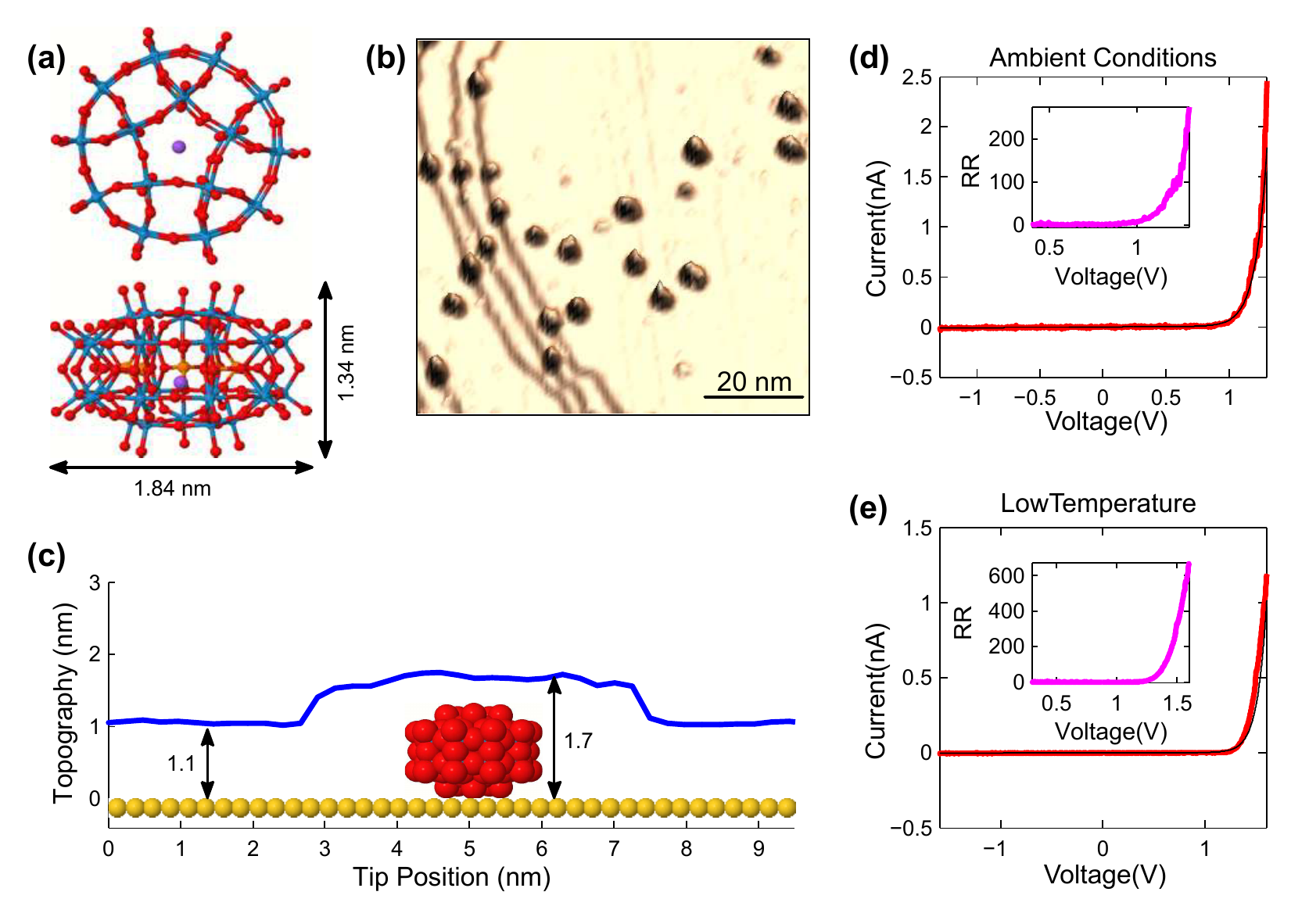}

\caption{Tunneling spectroscopy on a single polyoxometalate (POM) molecule deposited on an atomically flat gold surface. (a), top and side view of the   $\rm [DyP_{5}W_{30}O_{110}]^{12-}$  polyoxometalate molecule (red: oxygen; blue: tungsten; orange: phosphor; violet: dysprosium) and dimensions considering the van der Waals radius for the oxygen atoms.  (b), Scanning tunneling microscopy image in ambient conditions of the POM molecule on an annealed Au substrate (Vbias = +1.2 V, applied to the substrate, I = 0.1 nA). Bias voltages in excess of +1.2 V are required for scanning to avoid sweeping the molecules with the tip (see supplementary Fig. S1). The molecules physisorb flat on terraces and steps of the Au(111)  surface and are found to be stable in air. (c), Topographic profile (blue line) showing the typical  scanning height  on the substrate ($\sim1.1$ nm) and on the molecule ($\sim1.7 $ nm) for the usual scanning conditions. (d), (e) Tunneling I-V curve (red curve) above the molecule at ambient conditions and at low temperatures (T = 350 mK), respectively. The thin black curve indicates the fit to \Eref{EqnCurrent1}. The corresponding insets show the rectification ratio   as a function of voltage. These curves are an average of 20 individual curves. }
\label{Fig1}
\end{center}

\end{figure}

Our proof-of-principle demonstration is based on the fabrication of single-molecule rectifiers using polyoxometalate (POM) clusters deposited on gold that were individually contacted with a scanning tunneling microscope (STM) tip both in ambient conditions and at low temperatures. POMs are electron-acceptor metal-oxide clusters with a rich variety of structures and electronic properties, which makes them very interesting in many nanotechnological applications \cite{Clemente}. In particular, $\rm [DyP_{5}W_{30}O_{110}]^{12-}$, the molecule used in this work, is a molecular polyoxoanion formed by a tungsten-oxide framework encapsulating a Dy ion in its internal cavity (see   \Fref{Fig1} (a)) \cite{Cardona}. In order to create metal-molecule-metal contacts, we first deposited this inorganic molecule on annealed gold. Imaging the molecules with STM shows isolated molecules on terraces and steps of the Au (111) substrate (see \Fref{Fig1}(b)). We target a single molecule and acquire I-V curves by placing the tip above the molecule. Between -1.3 V and +1.3 V, these I-V curves show a strongly rectifying behavior with a negligible current for negative voltages and an exponentially increasing current for positive voltages higher than one volt (See  \Fref{Fig1} (d) and (e)). Indeed the current can be accurately fitted by the expression

\begin{equation}
I= I_0 (e^{\frac{V}{V_T}}-1) + I_D (V)
\label{EqnCurrent1}
\end{equation}
where the first term has the form of the ideal diode equation with saturation current $I_0$. In the case of semiconducting diodes $V_T$, the so-called thermal voltage, depends linearly on temperature, in contrast, we observe no temperature dependence in our molecular diodes, finding  $V_T=0.11\pm0.02$ V both for ambient and liquid helium temperatures. The second term has the form $I_D=AV+BV^3$, where A and B are constants, corresponding to tunneling between two metals. As shown in the insets of \ref{Fig1} (d) and (e), the rectification ratio, defined as the magnitude of the ratio of the current for positive and negative voltages, RR(V)=I(V)/I(-V), for large voltages depends exponentially on voltage, as in the case of semiconductor diodes, and reaches very large values (over 250 in ambient conditions and over 600 at low temperatures). These rectification ratios reported here are more than ten times larger than those previously reported in single-molecule junctions(RR$\approx$11) \cite{Diez} and even larger than those observed in junctions based on molecular monolayers (RR$\approx$150)\cite{Ashwell, Nijhuis}.

\begin{figure}
\begin{center}
\includegraphics{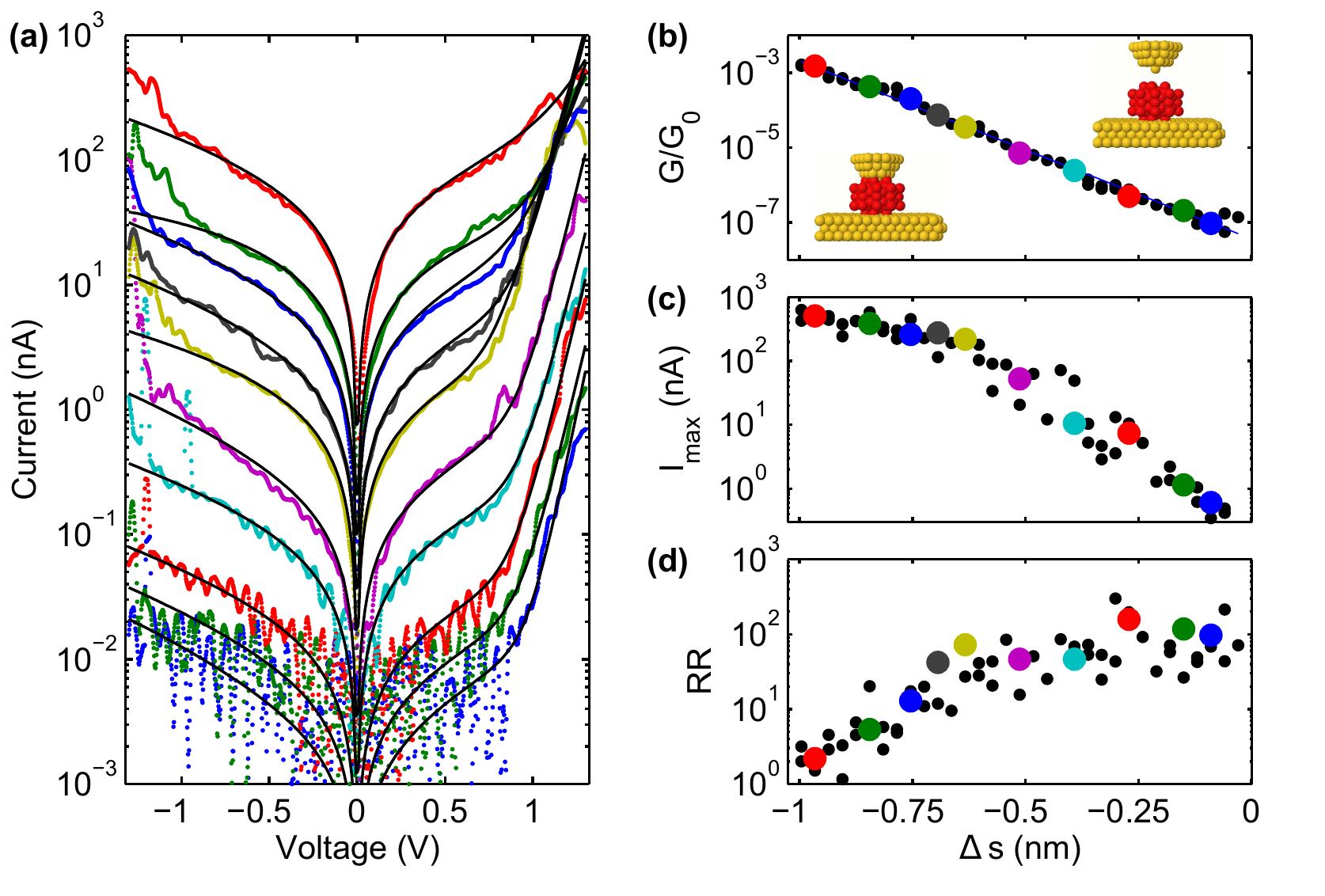}
\caption{Tip-molecule separation dependence of rectification in a single molecule junction. (a) Selection of the I-V  characteristics acquired as tip-molecule separation s is reduced (linear-log scale). The thin black lines are a fit to the diode model described by \Eref{EqnCurrent}. Each curve has been acquired in about 100 ms. (b), Low bias conductance (in log scale) as the tip-molecule separation decreases. (c), (d) Maximal current  $I_{max}$  at V = +1.3 V. and rectification ratio, RR, respectively. The black points in (b), (c), (d) correspond to the positions where I-V curves have been acquired, the coloured points correspond to the I-V curves in (a).}
\label{Fig2}
\end{center}

\end{figure}

Next we explore the dependence of the I-V curves on the tip height over the molecule by approaching the tip to the molecule in 30 pm steps and acquiring I-V curves at each step (see \Fref{Fig2} (a)). As the tip-molecule separation $ s $ decreases, we observe that for changes in separation in the range $ -0.5 \ \rm{nm} <\Delta s < 0 \ \rm{nm} $, the I-V curves are almost identical, scaling with a factor $e^{-\beta \Delta s}$, with $\beta \approx 11 $ \rm{nm}$^{-1}$. This exponent $\beta$ is similar to that corresponding to tunneling on the bare gold substrate in ambient conditions (see supplementary Figure S2). Further reduction in tip-molecule separation results in the saturation of the maximal current leading to an exponential decrease in the rectification ratio RR (\Fref{Fig2} (c) and (d)). This change is related to tip-molecule contact formation, and is consistent with the tip height estimated from the apparent height of the molecule on the gold substrate (see \Fref{Fig1} (c)). We have studied over 20 different molecular junctions all of them presenting a very similar behavior to the one just described.

For a functional diode not only a high rectification ratio is needed but also a large enough current. In \Fref{Fig2} (c) and (d), we observe that currents higher than 100 nA are possible with rectification ratios higher than 100. These currents correspond to enormous current densities of up to $10^7 \rm{A/cm^2}$, which are many orders of magnitude larger than those reported for planar molecular junctions \cite{Nijhuis}, which range from $10^{-2}$ to 1 A/cm$^2$.

\begin{figure}
\begin{center}
\includegraphics{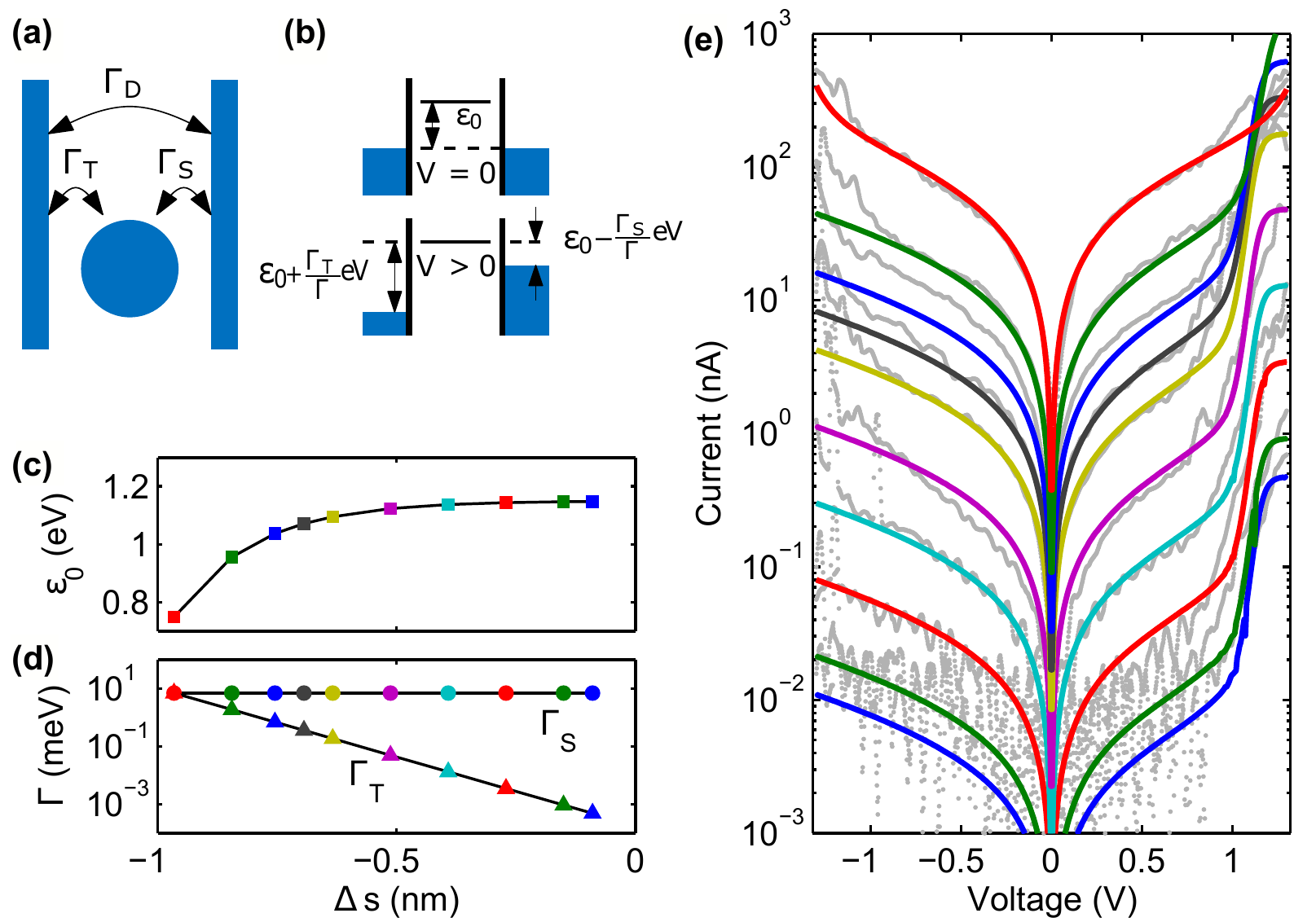}
\caption{One-level model with the addition of a direct channel. (a) The molecule is coupled to the substrate and to the tip through tunneling rates $\Gamma_S$, and $\Gamma_T$, receptively, and additionally tip and substrate are coupled directly through tunneling rate $\Gamma_D$. (b) Diagram for voltage drop. The position of the level with respect to the Fermi level of the electrodes depends on the applied voltage and on the tunneling rates $\Gamma_S$, and $\Gamma_T$.  (c), (d) Position of energy level, $\varepsilon_0$, and tunneling rates $\Gamma_S$, (circles) and $\Gamma_T$, (triangles), as a function of tip position. (e), I-V curves obtained from the one-level model (coloured curves) superposed on the experimental curves (gray curves). The coloured symbols in (c), (d) correspond to the coloured curves in (e).}
\label{Fig3}
\end{center}

\end{figure}

The charge transport process through the molecular junction can be explained using a simple model that clarifies the relevant physical processes giving insight into the mechanism for rectification. This model, which is sketched in \Fref{Fig3}(a), considers that the total current through the junction, $I=I_D+I_M$, is given by the contribution of two different conduction paths or channels: a direct channel (from tip to substrate), giving rise to a current $I_D$, and a molecular channel involving the electronic structure of the molecule, contributing a current $I_M$. 
The direct channel involves coupling between the tip and the substrate, and can be described via an energy-independent tunneling rate 
$\Gamma_D = \pi t_D^2 \rho(E_F)$, where $t_D$ is the matrix element describing the coupling between the two electrodes and $\rho(E_F)$ is the local density of states at the electrodes, which are assumed to be identical  \cite{Cuevas}. This direct channel accounts for the tunnel-like current $I_D$, whose linear part can be written as $I_D= \frac{2e^2}{h} 4 \pi \Gamma_D \rho(E_F) V$. The molecular channel can be described in terms of a single-level model \cite{Larade}. In this model one assumes the charge transport to be dominated by a single molecular level, the position of which in the absence of bias is denoted by $\varepsilon_0$  measured with respect to the Fermi energy. This level corresponds to the lowest unoccupied molecular orbital or LUMO, for $\varepsilon_0 > 0$   or to the highest occupied molecular level or HOMO, for  $\varepsilon_0 < 0$   and is coupled to the metallic electrodes via energy-independent tunneling rates $\Gamma_S$   and $\Gamma_T$, which describe the strength of the coupling of the molecule to the substrate and to the STM tip, respectively. These tunneling rates can be written as $\Gamma_S = \pi t_S^2  \rho (E_F)$ and $\Gamma_T = \pi t_T^2  \rho (E_F)$ , where $t_{T}$ ($t_S$) is the matrix element describing the coupling between the tip (substrate) and the molecule \cite{Cuevas}. In the spirit of the Landauer formalism \cite{Larade}, the current in this model is given by

\begin{equation}
I_M (V)= \frac{2e}{h}\int_{-\infty}^{+\infty} \ T(E,V) \ [f(E - eV /2) - f(E+ eV /2)] \ dE 
\label{EqnCurrent}
\end{equation}
where f(E)  is the Fermi function, V  is the voltage of the substrate with respect to the tip, and  T(E,V)  is the energy- and voltage-dependent transmission that is expressed as

\begin{equation}
T(E,V) = \frac{4 \Gamma_S \Gamma_T} {[E-\varepsilon_0(V)]^2 + \Gamma^2}
\label{EqnTrans}
\end{equation}
Here,  $\Gamma=\Gamma_S+\Gamma_T$ and  $\varepsilon_0(V)$ is the level position that we assume to shift with bias as $\varepsilon_0(V) = \varepsilon_0 + (eV/2)[(\Gamma_L - \Gamma_R)/\Gamma]$ . This bias dependence takes into account the fact that a larger portion of the voltage must drop at the interface with the weaker coupling. It is precisely this asymmetric voltage profile, which appears when $\Gamma_S \neq \Gamma_T$, the origin of the rectifying behavior. When the STM tip is above the molecule, $\Gamma_S>\Gamma_T$, the onset of the current will occur at  $eV= \varepsilon_0 \Gamma / \Gamma_S$  , a positive onset like that shown in \Fref{Fig1} (d) and (e) indicates that the molecular level is the LUMO. This is consistent with the electron acceptor character of the POM cluster \cite{Pope}.

To model the variation in the I-V characteristics as the tip approaches the molecule, we assume that the tip-molecule coupling, $\Gamma_T$  and the tip-substrate coupling,  $\Gamma_D$ have an exponential dependence on tip-molecule separation, $\Gamma_D, \ \Gamma_T \sim e^{-\beta \  \Delta s}$, with $\beta \approx 11 \ nm^{-1}$, as expected for tunneling, while the molecule-substrate coupling, $\Gamma_S$ remains constant. As a good tip-molecule contact is formed both couplings to the molecule will become identical, i.e., $\Gamma_S=\Gamma_T$  . We also allow for a shift in the position of the molecular level $\varepsilon_0$ , which may occur as a consequence of charge transfer between the molecule and the tip. In  \Fref{Fig3} (e) we show the evolution of the I-V curves according to this model, with values for $\varepsilon_0$,  $\Gamma_S$ and $\Gamma_T$ given in \Fref{Fig3} (c) and (d).  In spite of the simplicity of the model that does not take into consideration level degeneracy, proximity of several molecular levels, etc., we find a good semi-quantitative agreement with the experiment.

\begin{figure}
\begin{center}
\includegraphics{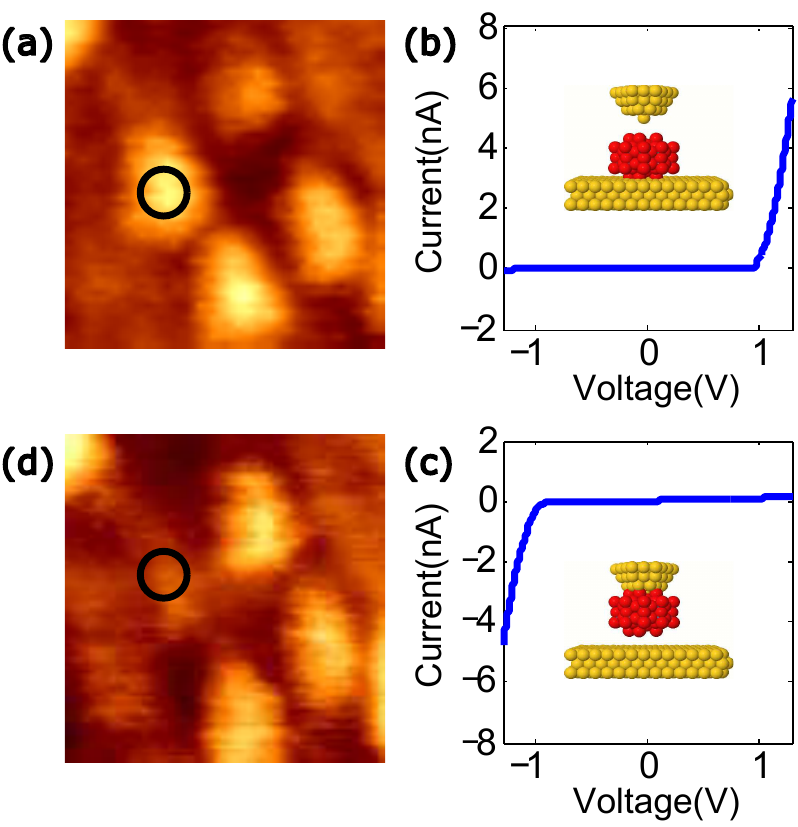}
\caption{Inversion of rectification as the molecule is transferred from the substrate to the tip. (a), STM image before approach experiment. (b) Tunneling I-V characteristic on the molecule in the center of the red circle. (c) Tunneling I-V characteristic after performing an approach experiment. The inversion of the I-V characteristic is due to transfer of the molecule to the tip. (d), STM image after approach experiment. The target molecule has been transferred to the tip and then lost while scanning.}
\label{Fig4}
\end{center}
\end{figure}

The validity of this model is further supported by the reversal of rectification observed when the molecule becomes attached to the tip (see \Fref{Fig4}). Indeed further pressing on the molecule may result in partial burying into the gold substrate (see supplementary Figure S5) or more interestingly in attachment of the molecule to the tip. In this case, according to the model, the current onset will occur at  $eV= -\varepsilon_0 \Gamma / \Gamma_T$, explaining the reversal of rectification observed in \Fref{Fig4} (c).

\section{Conclusions}

From the analysis above, it is clear that the rectification mechanism described in this work requires two ingredients: (1) An asymmetry in the coupling of the molecule to the metal electrodes, i.e., $\Gamma_T<\Gamma_S$, and (2) an asymmetry in the electronic structure of the molecule, with either the HOMO or the LUMO located much closer in energy to the Fermi level of the electrodes than the other. We can also identify the conditions for the simultaneous optimization of rectification and current density. In order to have a steep current onset the coupling of the molecule to the substrate must be small to avoid level broadening, i.e., $\Gamma_S<< |\varepsilon_0|$. On the other hand, the coupling to the tip $\Gamma_T$  should be smaller than the coupling to the substrate but not be too small since that would reduce the current. Thus, ideally a molecular diode should work in the regime $\Gamma_T< \Gamma_S<< |\varepsilon_0|$, with $\Gamma_T$ as large as possible. This implies that the metal-molecule coupling must be \textit{moderately }asymmetric. In the present case, this coupling is expected to take place between the LUMO of the POM and the metallic substrate. Since the LUMO is essentially localized on the metal ions \cite{Fernandez} its coupling with the substrate is expected to be relatively low, thus preventing level broadening. This ideal situation contrasts with that of fullerene  derivatives, as these are $\pi$-conjugated molecules with accessible p-orbitals at their surfaces, which favors a larger coupling with the substrate \cite{Zhao1}.

To conclude, we must remark that this mechanism for rectification is by no means restricted to a particular type of molecule, and could be implemented in both organic and inorganic molecules, providing a simple strategy for the rational design of molecular diodes. Finally there is no fundamental obstacle to scale up our molecular diodes without affecting the rectification ratio by introducing eventually a monolayer of these molecules.

\section*{Acknowledgment}
This work was supported by Spanish MICINN/MINECO through the programs MAT2011-25046, MAT2011-22785, FIS2011-28851-C02-01 and MAT2014-57915-R; Comunidad de Madrid through programs NANOFRONTMAG-CM (S2013/MIT-2850) and MAD2D-CM (S2013/MIT-3007),  the European Union (FP7) through programs ITN “MOLESCO” Project Number 606728, Project ELFOS, and ERC Advanced Grant SPINMOL; and the Generalitat Valenciana, through programs PROMETEO and ISIC Nano.

%
%


\clearpage 
\noappendix

\end{document}